\documentclass[prb, aps, superbib, tightenlines, twocolumn, floatfix,
superscriptaddress]{revtex4}
\usepackage{graphicx, dcolumn, bm}

\renewcommand{\vec}{\bf}
\begin{document}

\title{Electron turbulence at nanoscale junctions}
\author{Neil Bushong}
\email{bushong@physics.ucsd.edu}
\affiliation{
Department of Physics, University of California, San Diego, La Jolla,
CA 92093-0319}

\author{John Gamble}
\affiliation{
Department of Physics, The College of Wooster, Wooster, Ohio
44691-2363}

\author{Massimiliano {Di Ventra}}
\affiliation{
Department of Physics, University of California, San Diego, La Jolla,
CA 92093-0319}

\date{\today}

\begin{abstract}
Electron transport through a nanostructure can be characterized in
part using concepts from classical fluid dynamics.  It is thus natural
to ask how far the analogy can be taken, and whether the electron
liquid can exhibit nonlinear dynamical effects such as turbulence.
Here we present an \emph{ab-initio} study of the electron dynamics in
nanojunctions which reveals that the latter indeed exhibits behavior
quite similar to that of a classical fluid.  In particular, we find
that a transition from laminar to turbulent flow occurs with
increasing current, corresponding to increasing Reynolds
numbers. These results reveal unexpected features of electron dynamics
and shed new light on our understanding of transport properties of
nanoscale systems.
\end{abstract}

\maketitle

Electron transport through a nanoscale junction is usually described
as a scattering problem.\cite{landauer57, buttiker85, diventra00,
taylor01, damle01, palacios02} On the other hand, it has been
shown\cite{martin59, vignale97, tokatly05, dagosta06jpcm} that the
behavior of the electron liquid obeys dynamical equations of motion
which are similar in form to those governing the dynamics of classical
liquids. In particular, it was recently shown\cite{dagosta06jpcm} that
the time-dependent Schr{\"o}dinger equation (TDSE) for electrons
flowing across a nanostructure can be cast in the form of generalized
Navier-Stokes equations. A consequence of this analogy is the
prediction that under certain conditions the laminar flow of the
liquid may become unstable and turbulent behavior is
expected.\cite{dagosta06jpcm, frisch:turbulence, landau:fluid} One can
then borrow knowledge from classical fluid dynamics and hypothesize
that the electron flow will make a transition from laminar to
turbulent regimes, if, e.g., the current is increased. Like in a
classical liquid, if, for instance, electrons flow from one electrode
(call it the top electrode) to another (call it the bottom electrode)
across a nanojunction, turbulence would first manifest itself with the
break-up of the top-down symmetry of the electron
flow.\cite{frisch:turbulence, landau:fluid} By increasing the current
further one should observe the formation of eddies in proximity to the
junction. These eddies are created because of the larger kinetic
energy in the direction of current flow (longitudinal kinetic energy)
compared to the transverse direction (transverse kinetic energy). By
increasing the current further, the disparity between longitudinal and
transverse components of the kinetic energy increases, and the system
flow eventually breaks any remaining symmetry, thus fully developing
turbulence.\cite{frisch:turbulence, landau:fluid} Despite the
prediction of turbulent behavior of the electron liquid in
nanostructures\cite{dagosta06jpcm} an explicit demonstration of this
phenomenon and the analysis of its microscopic features has not been
presented yet.

In this communication we set to show numerically turbulent effects for
the electron liquid crossing a nanojunction both by solving directly
the TDSE, and by solving the generalized Navier-Stokes equations
derived in Ref.~\onlinecite{dagosta06jpcm}. These read
\begin{eqnarray}
D_t n({\vec r}, t) = & - n({\vec r}, t)
    {\vec \nabla} \cdot {\vec u}({\vec r}, t) \, , \nonumber \\
m n({\vec r}, t) D_t u_i({\vec r}, t) = & - \partial_i P({\vec r},t)
    + \partial_j \pi_{i, j} \nonumber \\
    & - n({\vec r}, t) \partial_i V_{\rm ext}({\vec r}, t) \, .
    \label{eq:NS}
\end{eqnarray}
Here, $n$ is the electron density, ${\vec u}={\vec j}/n$ is the
velocity field, i.e. the ratio between the current density ${\vec j}$
and the density, and $D_t=\frac{\partial}{\partial t}+{\bf u}\cdot
\nabla$ is the convective derivative.\cite{landau:fluid} $P({\vec r},
t)$ is the pressure of the liquid, $V_{\rm ext}({\vec r}, t)$ an
external potential, and $\pi_{i,j}$ is a traceless tensor that
describes the shear effect on the liquid. It has the form
\begin{equation}
\pi_{i,j} = \eta \left( \partial_i u_j + \partial_j u_i
    - \frac{2}{3} \delta_{i,j} \partial_k u_k \right) \, ,
\end{equation}
where $\eta=\hbar n f(n)$ is the viscosity of the electron liquid
and $f(n)$ is a smooth function of the density.\cite{conti99}

\begin{figure*}
\includegraphics[width=6.9in]{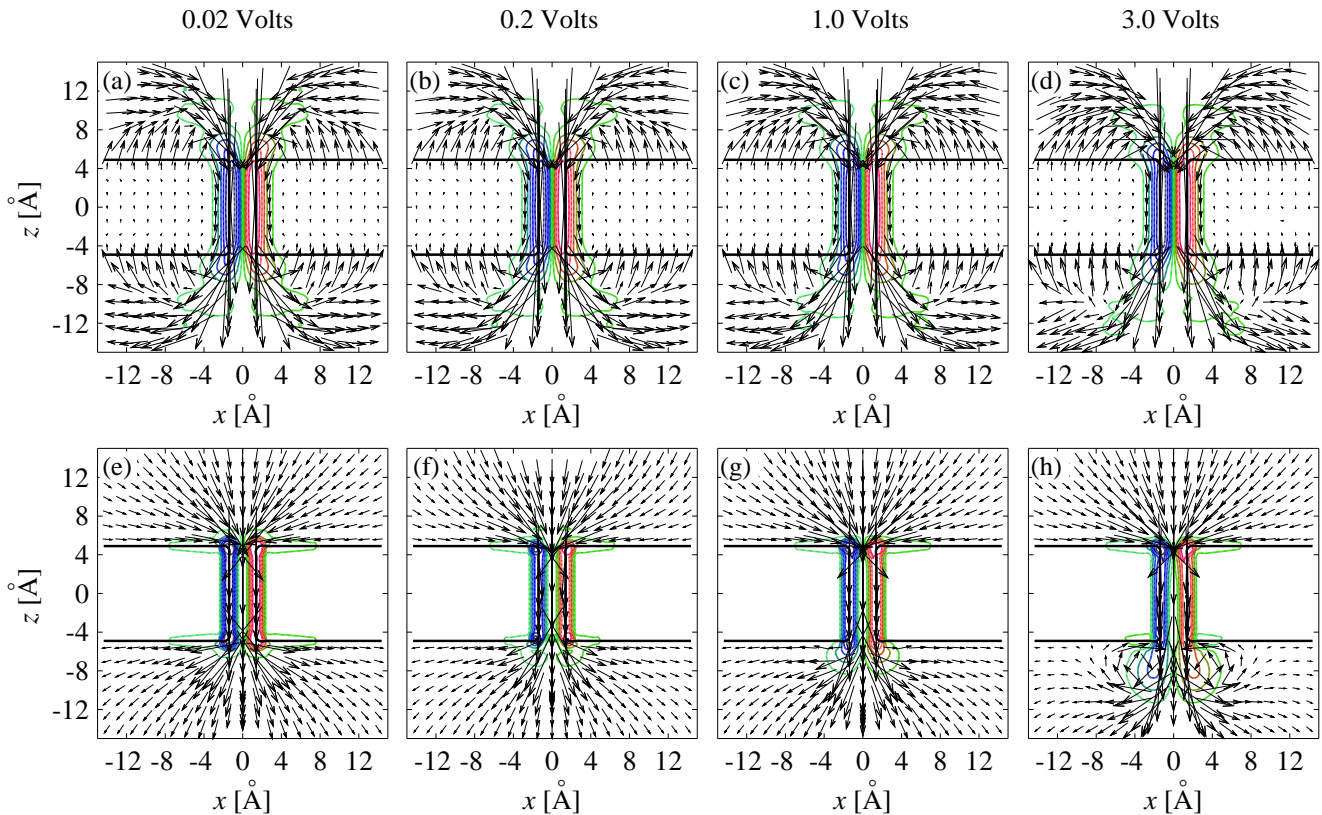}
\caption{(Color online) Panels (a)-(d): Electron current density for
electrons moving from the top electrode to the bottom electrode across
a nanojunction at $t = 1.4$ fs, for an initial bias of (a) 0.02 V, (b)
0.2 V, (c) 1.0 V and (d) 3.0 V. The arrows denote the current density,
while the level sets denote the curl of the 2D current density. The
solid lines delimit the contour of the junction. Panels (e)-(h):
Velocity field solution of the equations~(\ref{eq:NS}), for a liquid
with same velocity, density and viscosity as the quantum mechanical
one.} \label{fig:curls}
\end{figure*}

In analogy with the classical case we expect that the atomic
structure, and in particular atomic defects in proximity to the
junction, play the role of ``obstacles'' for the liquid and thus favor
turbulence. Since we aim at showing that turbulence develops
irrespective of the underlying atomic structure, we consider electrons
interacting with a uniform positive background charge (i.e., the
``jellium'' model\cite{diventra02}). The system we consider therefore
consists of two large but finite jellium electrodes-- subject to a
bias-- connected via a nanoscale jellium bridge.  (The jellium edge of
this system is represented with solid lines in each panel of
Fig.~\ref{fig:curls}.)  We choose the density of the jellium at
equilibrium typical of bulk gold ($r_s\approx3a_0$). For computational
convenience we choose a quasi-2D system, approximately 2.8
\AA~thick.\footnote{Each electrode is 51.8 \AA~wide in the $x$
direction, and 22.4 \AA~long in the $z$ direction of current flow (see
Fig.~\ref{fig:curls}). The width of the rectangular bridge is 2.8 \AA,
and the gap between the electrodes is 9.8 \AA.}  Note that, everything
else being equal, a quasi-2D geometry {\em disfavors} turbulence
compared to a 3D one.  We thus expect that if turbulence develops in
our chosen quasi-2D geometry with given thickness, then turbulence
will develop even more easily if we leave everything else unchanged
(including the total current) and increase the thickness of the
electrodes.

The solution of the TDSE for the many-body system is obtained within
Time-Dependent Current Density Functional
Theory(TDCDFT);\cite{vignale97} i.e., for each single-particle state
$\phi_\alpha$, we have solved the equation of motion (in atomic units)
\begin{equation}
\left\{ i \frac{\partial}{\partial t}
    - \frac{1}{2} \left( \frac{1}{i}{\vec \nabla} - \frac{1}{c}
    {\vec a}_{\rm xc} \right)^2 - v_{\rm jel} - v_{\rm H} - v_{\rm
    xc}
    \right\} \phi_\alpha = 0
    \label{eq:tdse}
\end{equation}
where $c$ is the speed of light, $v_{\rm jel}$ is the potential due to
the jellium, $v_{\rm H}$ is the Hartree potential, and $v_{\rm xc}$ is
the exchange-correlation scalar potential.\footnote{We have used the
adiabatic local density approximation to the scalar
exchange-correlation potential,\cite{kohn65, zangwill80, runge84} as
derived by Ceperley and Alder\cite{ceperley80} and parametrized by
Perdew and Zunger.\cite{perdew81}} The shear viscosity of the electron
liquid $\eta$ enters the problem through the exchange-correlation
vector potential ${\vec a}_{\rm xc}$.\cite{vignale97, conti99} If we
make the appoximation that, at any given time, the viscosity is a
function of the density only, and does not depend on time explicitly,
we find that the $i^{\rm th}$ component of ${\vec a}_{\rm xc}$ evolves
according to\cite{conti99}
\begin{eqnarray}
\lefteqn{
    \frac{1}{c} \frac{\partial a_{{\rm xc}, i}}{\partial t }
    = \frac{1}{n} \sum_j \bigg\{
        \eta \frac{\partial^2 u_i}{\partial r_j^2}
    + \frac{\eta}{3} \frac{\partial^2 v_j}{\partial_i \partial_j}
} \nonumber \\
& & + \frac{\partial \eta}{\partial r_j}
    \bigg( \frac{\partial u_i}{\partial r_j}
+ \frac{\partial u_j}{\partial r_i} \bigg) - \frac{2}{3}
\frac{\partial \eta}{\partial r_i}
    \frac{\partial u_j}{\partial r_j}
\bigg\} \, .
\end{eqnarray}

For the viscosity $\eta$ of the electron liquid we have used the one
reported in Ref.~\onlinecite{conti99}.  We employ the approach
described in Refs.~\onlinecite{diventra04, bushong05, sai07, cheng07}
to initiate electron dynamics and calculate the current. We prepare
the system by placing it in its ground state; because of the applied
bias, the system exhibits a separation of charge.  At a time $t=0$, we
remove the bias, and let the system evolve according to
eq.~(\ref{eq:tdse}).\footnote{The grid spacing of the jellium system
is 0.7 \AA, and the timestep used to propagate the system $2.5 \times
10^{-3}$ fs. We used the Chebyshev method for constructing the
time-evolution operator.\cite{tal-ezer84}} After long time scales, the
electrons encounter the far boundaries of the jellium electrodes and
reflect. However, we are interested in comparing the electron dynamics
calculated from eq.~(\ref{eq:tdse}) with the one obtained by solving
the Navier-Stokes equations~(\ref{eq:NS}), at times smaller than this
reflection time. Equations~(\ref{eq:NS}) are solved by assuming the
same density, initial velocity and viscosity of the liquid employed in
the TDCDFT calculation.\footnote{For the simulation of the
Navier-Stokes equations we use Dirchlet boundary conditions for the
velocity at the inlet, and Neumann boundary conditions at the outlet.}
We also solve eqs.~(\ref{eq:NS}) assuming the liquid to be
incompressible. This simplifies the calculations enormously but leads
to some minor differences with the solutions of eq.~(\ref{eq:tdse})
(see discussion later).

Fig.~\ref{fig:curls} depicts the flow of electrons across the
nanostructure, for a range of biases between 0.02V and 3.0V, after the
initial transient. The panels (a)-(d) correspond to the solution of
eq.~(\ref{eq:tdse}); the panels (e)-(h) to the solution of the
eqs.~(\ref{eq:NS}) using the same set of parameters.\cite{gerris}
Panel (a) has to be compared with panel (e); panel (b) with (f), and
so on. As anticipated, we observe some differences between the
solutions of the eqs.~(\ref{eq:NS}) and the solutions of the
eq.~(\ref{eq:tdse}).  These differences are due to the details of the
charge configuration at the electrode-junction interface, and some
degree of compressibility of the quantum liquid in the junction.

From Fig.~\ref{fig:curls} we can see the effect of surface charges, in
that some electrons flow parallel to the surfaces.\cite{sai07} More
importantly, at low biases, the flow is laminar and ``smooth''.  In
addition, at these biases the current density shows an almost perfect
top-bottom symmetry: the direction of the flow is symmetric with
respect to the operation $z\rightarrow -z$. This symmetry is even more
evident by comparing the curl of the current density in the top and
bottom electrodes (see for instance Fig.~\ref{fig:curls}(a) and (e)).

By increasing the bias, however, a transition occurs: the symmetry
$z\rightarrow -z$ of the current density breaks completely, and eddies
start to appear in proximity to the junction. This is clearly evident,
for instance, in Fig.~\ref{fig:curls}(d) and (h). The outgoing current
density in the bottom electrode has a more varied angular behavior, in
contrast to the behavior in the top electrode, in which the electron
liquid flows more uniformly toward the junction.

Since the panels (e)-(h) of Fig.~\ref{fig:curls} practically describe
the dynamics of a classical fluid with the same parameters as the
quantum liquid, the analogy between the electron flow and the one of a
classical liquid is quite evident. We can push this analogy even
further by defining a Reynolds number for the quantum system as well:
$R = u_zL\rho / \eta$, where $u_z$ is the longitudinal velocity in the
junction, $L$ is the width of the junction, and $\rho$ is the density.
Using the density of valence electrons in gold, and using the current
density in the junction at $t=1.4$ fs, we obtain for the quantum case
the following Reynolds numbers: 0.216, 2.16, 10.8 and 32.5 for 0.02 V,
0.2 V, 1.0 V, and 3.0 V, respectively.

Just like in the classical case, we can then reinterpret the above
results as follows. At low Reynolds numbers, the flow is highly
symmetric from top to bottom.  This symmetry is lost as the Reynolds
number is increased.  At high Reynolds numbers, the incident flow is
laminar, while the outgoing flow has a jet-like character, and ``turns
back'' on itself creating local eddies in the current density.

\begin{figure}
\includegraphics[width=3.4in]{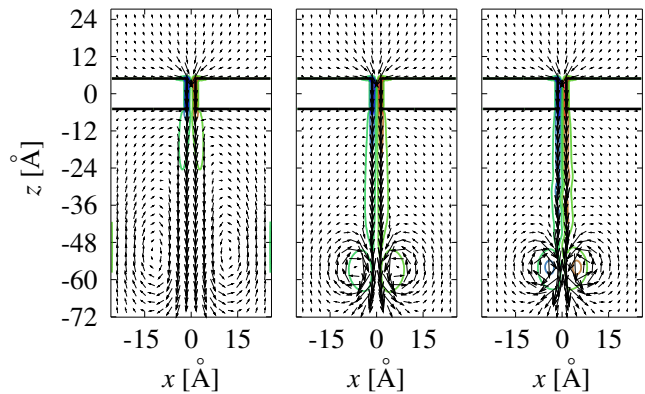}
\caption{(Color online) Current density (arrows) and curl of the
current density (denoted by level sets) of the electron liquid, for
three different Reynolds numbers, 32.5 (left panel), 162 (middle
panel), and 325 (right panel). Note that the fluid velocity has lost
perfect left-right symmetry in the middle- and right-panel cases.}
\label{fig:long_time}
\end{figure}

Having shown the similarity between the current flow obtained using
the Navier-Stokes equations~(\ref{eq:NS}) and the one obtained solving
eq.~(\ref{eq:tdse}) we can study the first one at times scales
prohibitive for full quantum mechanical simulations.  We can also
study the effect of a larger thickness of the electrodes on the
turbulent flow by realizing that this is equivalent to increasing the
Reynolds number (which, incidentally, is also equivalent to increasing
the bias).  This is illustrated in Fig.~\ref{fig:long_time} where the
current density and the curl of the current density are plotted for
the Reynolds number 32.5 of Fig.~\ref{fig:curls}(h) (left panel of
Fig.~\ref{fig:long_time}); same system but with a Reynolds number five
times larger (middle panel of Fig.~\ref{fig:long_time}); and (right
panel of Fig.~\ref{fig:long_time}) with a Reynolds number ten times
larger.

From Fig.~\ref{fig:long_time} it is evident that by increasing
thickness the last remaining symmetry $x\rightarrow -x$ is broken at
earlier times, leading to turbulent behavior closer to the
junction. For instance, in the case represented in
Fig.~\ref{fig:long_time} (middle panel), the left-right symmetry is
lost at about 14 fs, with consequent asymmetric flow within about 50
\AA$\;$ from the junction center. For the structure represented in
Fig.~\ref{fig:long_time} (right panel), the symmetry is broken at
about 6 fs, and the flow asymmetry appears at about 25 \AA$\;$ from
the junction.

\begin{figure}
\includegraphics[width=3.4in]{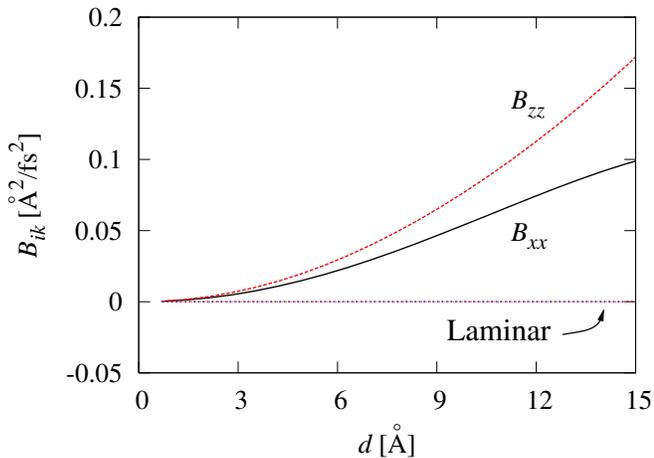}
\caption{(Color online) Various components of the velocity correlation
tensor $B_{ik}$ as a function of distance $d = |\delta{\vec r}|$, for
a Reynolds number of $R = 32.5$.  For the case where $R = 0.216$, the
elements of $B_{ik}$ are orders of magnitude smaller, and so the
corresponding curves for the laminar case coincide with the x axis.}
\label{fig:corr_fn}
\end{figure}

We can better quantify the amount of turbulence by calculating the
velocity correlation tensor\cite{landau:fluid}
\begin{equation}
B_{ik} = \langle (v_i({\vec r}) - v_i({\vec r} + \delta{\vec r}))
    (v_k({\vec r}) - v_k({\vec r} + \delta{\vec r})) \rangle \, ,
\end{equation}
where $\delta{\vec r}$ is a given distance, and $i,k=x,y,z$.  Here,
the angle brackets $\langle \cdots \rangle$ denote averaging over all
positions ${\vec r}$ within a given region. Fully developed and
isotropic turbulence has a velocity correlation tensor that is a
function only of the magnitude of $\delta{\vec r}$, and increases
quadratically with distance.\cite{landau:fluid} Instead, the
turbulence in the examples of Fig.~\ref{fig:long_time} is not fully
developed. The velocity correlation tensor, $B_{ik}$, thus depends on
both the magnitude of $\delta{\vec r}$ as well as its direction.  This
is illustrated in Fig.~\ref{fig:corr_fn} where various components of
$B_{ik}$ are plotted at $t=75.0$ fs for the system with Reynolds
number 32.5 as a function of the magnitude of $\delta{\vec r}$, where
we have chosen $\delta{\vec r}$ to point in the longitudinal (z)
direction. The spatial averaging has been carried out over the
left-hand side of the outgoing region (that is, in the region $z$ =
[-27.3 \AA, -72.1 \AA], $x$ = [-25.9 \AA, 0.0 \AA], where the origin
is in the center of the junction). For comparison, the same quantity
is plotted for the laminar case, i.e. for a Reynolds number of
.216. (To compare the laminar and turbulent cases we have scaled the
average turbulent velocity to the average laminar velocity.) As
expected, in the laminar case the correlation tensor is essentially
zero, while for the turbulent case it increases with distance.

We conclude by noting that an experiment in which the electron flow
can be monitored directly may measure nonlaminar electron behavior as
an asymmetry between the incoming and outgoing patterns of the current
density through a nanojunction. Experiments similar to the ones
reported in Ref.~\onlinecite{topinka00}, which use scanning probe
microscopy to image the flowlines, may provide such
capabilities. Note, however, that since these scanning probe
techniques record images over time scales much longer than the
turbulent electron dynamics, one would expect that in the turbulent
regime images of current flow appear ``smeared out'' compared to the
laminar case, and asymmetric in the incoming and outgoing patterns.
From the present work and the analytical results obtained in
Ref.~\onlinecite{dagosta06jpcm}, we also suggest that fully 3D and
non-adiabatic junctions, i.e. junctions with a geometry that changes
abruptly (like the one explored in this work), are the best candidates
to observe turbulence. We also expect defects and other impurities to
favor turbulent behavior by playing the role of ``obstacles'' for the
electron flow.

We finally note that turbulent behavior may have consequences on the
formation (or lack thereof) of local equilibrium distributions in the
electrodes,\cite{bushong05} and may generate non-trivial local
electron heating effects in the junction\cite{dagosta06nl} and at the
eddies sites; phenomena and properties which are still poorly
understood at the nanoscale and ultimately may have unexpected
consequences on the stability of nanostructures under current
flow.\cite{huang06} \\

We acknowledge useful discussions with Roberto D'Agosta.  One of us
(JG) received support from the National Science Foundation's REU
program.  This work was supported by the U.S.  Department of Energy
under grant DE-FG02-05ER46204.

\bibliography{turbulence}

\begin{thebibliography}{28}
\expandafter\ifx\csname natexlab\endcsname\relax\def\natexlab#1{#1}\fi
\expandafter\ifx\csname bibnamefont\endcsname\relax
  \def\bibnamefont#1{#1}\fi
\expandafter\ifx\csname bibfnamefont\endcsname\relax
  \def\bibfnamefont#1{#1}\fi
\expandafter\ifx\csname citenamefont\endcsname\relax
  \def\citenamefont#1{#1}\fi
\expandafter\ifx\csname url\endcsname\relax
  \def\url#1{\texttt{#1}}\fi
\expandafter\ifx\csname urlprefix\endcsname\relax\def\urlprefix{URL }\fi
\providecommand{\bibinfo}[2]{#2}
\providecommand{\eprint}[2][]{\url{#2}}

\bibitem[{\citenamefont{Landauer}(1957)}]{landauer57}
\bibinfo{author}{\bibfnamefont{R.}~\bibnamefont{Landauer}},
  \bibinfo{journal}{IBM J. Res. Dev.} \textbf{\bibinfo{volume}{1}},
  \bibinfo{pages}{223} (\bibinfo{year}{1957}).

\bibitem[{\citenamefont{B{\"u}ttiker et~al.}(1985)\citenamefont{B{\"u}ttiker,
  Imry, Landauer, and Pinhas}}]{buttiker85}
\bibinfo{author}{\bibfnamefont{M.}~\bibnamefont{B{\"u}ttiker}},
  \bibinfo{author}{\bibfnamefont{Y.}~\bibnamefont{Imry}},
  \bibinfo{author}{\bibfnamefont{R.}~\bibnamefont{Landauer}}, \bibnamefont{and}
  \bibinfo{author}{\bibfnamefont{S.}~\bibnamefont{Pinhas}},
  \bibinfo{journal}{Phys. Rev. B} \textbf{\bibinfo{volume}{31}},
  \bibinfo{pages}{6207} (\bibinfo{year}{1985}).

\bibitem[{\citenamefont{{Di Ventra} et~al.}(2000)\citenamefont{{Di Ventra},
  Pantelides, and Lang}}]{diventra00}
\bibinfo{author}{\bibfnamefont{M.}~\bibnamefont{{Di Ventra}}},
  \bibinfo{author}{\bibfnamefont{S.~T.} \bibnamefont{Pantelides}},
  \bibnamefont{and} \bibinfo{author}{\bibfnamefont{N.~D.} \bibnamefont{Lang}},
  \bibinfo{journal}{Phys. Rev. Lett.} \textbf{\bibinfo{volume}{84}},
  \bibinfo{pages}{979} (\bibinfo{year}{2000}).

\bibitem[{\citenamefont{Taylor et~al.}(2001)\citenamefont{Taylor, Guo, and
  Wang}}]{taylor01}
\bibinfo{author}{\bibfnamefont{J.}~\bibnamefont{Taylor}},
  \bibinfo{author}{\bibfnamefont{H.}~\bibnamefont{Guo}}, \bibnamefont{and}
  \bibinfo{author}{\bibfnamefont{J.}~\bibnamefont{Wang}},
  \bibinfo{journal}{Phys. Rev. B} \textbf{\bibinfo{volume}{63}},
  \bibinfo{pages}{245407} (\bibinfo{year}{2001}).

\bibitem[{\citenamefont{Damle et~al.}(2001)\citenamefont{Damle, Ghosh, and
  Datta}}]{damle01}
\bibinfo{author}{\bibfnamefont{P.~S.} \bibnamefont{Damle}},
  \bibinfo{author}{\bibfnamefont{A.~W.} \bibnamefont{Ghosh}}, \bibnamefont{and}
  \bibinfo{author}{\bibfnamefont{S.}~\bibnamefont{Datta}},
  \bibinfo{journal}{Phys. Rev. B} \textbf{\bibinfo{volume}{64}},
  \bibinfo{pages}{201403(R)} (\bibinfo{year}{2001}).

\bibitem[{\citenamefont{Palacios et~al.}(2002)\citenamefont{Palacios,
  P{\'e}rez-Jim{\'e}nez, Louis, San{F}abi{\'a}n, and Verg{\'e}s}}]{palacios02}
\bibinfo{author}{\bibfnamefont{J.~J.} \bibnamefont{Palacios}},
  \bibinfo{author}{\bibfnamefont{A.~J.} \bibnamefont{P{\'e}rez-Jim{\'e}nez}},
  \bibinfo{author}{\bibfnamefont{E.}~\bibnamefont{Louis}},
  \bibinfo{author}{\bibfnamefont{E.}~\bibnamefont{San{F}abi{\'a}n}},
  \bibnamefont{and} \bibinfo{author}{\bibfnamefont{J.~A.}
  \bibnamefont{Verg{\'e}s}}, \bibinfo{journal}{Phys. Rev. B}
  \textbf{\bibinfo{volume}{66}}, \bibinfo{pages}{035322}
  (\bibinfo{year}{2002}).

\bibitem[{\citenamefont{Martin and Schwinger}(1959)}]{martin59}
\bibinfo{author}{\bibfnamefont{P.~C.} \bibnamefont{Martin}} \bibnamefont{and}
  \bibinfo{author}{\bibfnamefont{J.}~\bibnamefont{Schwinger}},
  \bibinfo{journal}{Phys. Rev.} \textbf{\bibinfo{volume}{115}},
  \bibinfo{pages}{1342} (\bibinfo{year}{1959}).

\bibitem[{\citenamefont{Vignale et~al.}(1997)\citenamefont{Vignale, Ullrich,
  and Conti}}]{vignale97}
\bibinfo{author}{\bibfnamefont{G.}~\bibnamefont{Vignale}},
  \bibinfo{author}{\bibfnamefont{C.~A.} \bibnamefont{Ullrich}},
  \bibnamefont{and} \bibinfo{author}{\bibfnamefont{S.}~\bibnamefont{Conti}},
  \bibinfo{journal}{Phys. Rev. Lett.} \textbf{\bibinfo{volume}{79}},
  \bibinfo{pages}{4878} (\bibinfo{year}{1997}).

\bibitem[{\citenamefont{Tokatly}(2005)}]{tokatly05}
\bibinfo{author}{\bibfnamefont{I.~V.} \bibnamefont{Tokatly}},
  \bibinfo{journal}{Phys. Rev. B} \textbf{\bibinfo{volume}{71}},
  \bibinfo{pages}{165104} (\bibinfo{year}{2005}).

\bibitem[{\citenamefont{D'Agosta and {Di Ventra}}(2006)}]{dagosta06jpcm}
\bibinfo{author}{\bibfnamefont{R.}~\bibnamefont{D'Agosta}} \bibnamefont{and}
  \bibinfo{author}{\bibfnamefont{M.}~\bibnamefont{{Di Ventra}}},
  \bibinfo{journal}{J. Phys.: Condens. Matter} \textbf{\bibinfo{volume}{18}},
  \bibinfo{pages}{11059} (\bibinfo{year}{2006}).

\bibitem[{\citenamefont{Frisch}(1995)}]{frisch:turbulence}
\bibinfo{author}{\bibfnamefont{U.}~\bibnamefont{Frisch}},
  \emph{\bibinfo{title}{Turbulence}} (\bibinfo{publisher}{Cambridge University
  Press}, \bibinfo{year}{1995}).

\bibitem[{\citenamefont{Landau and Lifshitz}(1987)}]{landau:fluid}
\bibinfo{author}{\bibfnamefont{L.~D.} \bibnamefont{Landau}} \bibnamefont{and}
  \bibinfo{author}{\bibfnamefont{E.~M.} \bibnamefont{Lifshitz}},
  \emph{\bibinfo{title}{Fluid Mechanics}}, vol.~\bibinfo{volume}{6} of
  \emph{\bibinfo{series}{Course of Theoretical Physics}}
  (\bibinfo{publisher}{Pergamon Press, NY}, \bibinfo{year}{1987}),
  \bibinfo{edition}{2nd} ed.

\bibitem[{\citenamefont{Conti and Vignale}(1999)}]{conti99}
\bibinfo{author}{\bibfnamefont{S.}~\bibnamefont{Conti}} \bibnamefont{and}
  \bibinfo{author}{\bibfnamefont{G.}~\bibnamefont{Vignale}},
  \bibinfo{journal}{Phys. Rev. B} \textbf{\bibinfo{volume}{60}},
  \bibinfo{pages}{7966} (\bibinfo{year}{1999}).

\bibitem[{\citenamefont{{Di Ventra} and Lang}(2002)}]{diventra02}
\bibinfo{author}{\bibfnamefont{M.}~\bibnamefont{{Di Ventra}}} \bibnamefont{and}
  \bibinfo{author}{\bibfnamefont{N.~D.} \bibnamefont{Lang}},
  \bibinfo{journal}{Phys. Rev. B} \textbf{\bibinfo{volume}{65}},
  \bibinfo{pages}{045402} (\bibinfo{year}{2002}).

\bibitem[{\citenamefont{{Di Ventra} and Todorov}(2004)}]{diventra04}
\bibinfo{author}{\bibfnamefont{M.}~\bibnamefont{{Di Ventra}}} \bibnamefont{and}
  \bibinfo{author}{\bibfnamefont{T.~N.} \bibnamefont{Todorov}},
  \bibinfo{journal}{J. Phys. Cond. Matt.} \textbf{\bibinfo{volume}{16}},
  \bibinfo{pages}{8025} (\bibinfo{year}{2004}).

\bibitem[{\citenamefont{Bushong et~al.}(2005)\citenamefont{Bushong, Sai, and
  {Di Ventra}}}]{bushong05}
\bibinfo{author}{\bibfnamefont{N.}~\bibnamefont{Bushong}},
  \bibinfo{author}{\bibfnamefont{N.}~\bibnamefont{Sai}}, \bibnamefont{and}
  \bibinfo{author}{\bibfnamefont{M.}~\bibnamefont{{Di Ventra}}},
  \bibinfo{journal}{Nano Lett.} \textbf{\bibinfo{volume}{5}},
  \bibinfo{pages}{2569} (\bibinfo{year}{2005}).

\bibitem[{\citenamefont{Sai et~al.}()\citenamefont{Sai, Bushong, Hatcher, and
  {Di Ventra}}}]{sai07}
\bibinfo{author}{\bibfnamefont{N.}~\bibnamefont{Sai}},
  \bibinfo{author}{\bibfnamefont{N.}~\bibnamefont{Bushong}},
  \bibinfo{author}{\bibfnamefont{R.}~\bibnamefont{Hatcher}}, \bibnamefont{and}
  \bibinfo{author}{\bibfnamefont{M.}~\bibnamefont{{Di Ventra}}},
  \bibinfo{note}{{P}hys. Rev. B, in press}.

\bibitem[{\citenamefont{Cheng et~al.}(2006)\citenamefont{Cheng, Evans, and
  Voorhis}}]{cheng07}
\bibinfo{author}{\bibfnamefont{C.-L.} \bibnamefont{Cheng}},
  \bibinfo{author}{\bibfnamefont{J.~S.} \bibnamefont{Evans}}, \bibnamefont{and}
  \bibinfo{author}{\bibfnamefont{T.~V.} \bibnamefont{Voorhis}},
  \bibinfo{journal}{Phys. Rev. B} \textbf{\bibinfo{volume}{74}},
  \bibinfo{pages}{155112} (\bibinfo{year}{2006}).

\bibitem[{\citenamefont{Popinet}(2003)}]{gerris}
\bibinfo{author}{\bibfnamefont{S.}~\bibnamefont{Popinet}}, \bibinfo{journal}{J.
  Comput. Phys.} \textbf{\bibinfo{volume}{190}}, \bibinfo{pages}{572}
  (\bibinfo{year}{2003}).

\bibitem[{\citenamefont{Topinka et~al.}(2000)\citenamefont{Topinka, LeRoy,
  Shaw, Heller, Westervelt, Maranowski, and Gossard}}]{topinka00}
\bibinfo{author}{\bibfnamefont{M.~A.} \bibnamefont{Topinka}},
  \bibinfo{author}{\bibfnamefont{B.~J.} \bibnamefont{LeRoy}},
  \bibinfo{author}{\bibfnamefont{S.~E.~J.} \bibnamefont{Shaw}},
  \bibinfo{author}{\bibfnamefont{E.~J.} \bibnamefont{Heller}},
  \bibinfo{author}{\bibfnamefont{R.~M.} \bibnamefont{Westervelt}},
  \bibinfo{author}{\bibfnamefont{K.~D.} \bibnamefont{Maranowski}},
  \bibnamefont{and} \bibinfo{author}{\bibfnamefont{A.~C.}
  \bibnamefont{Gossard}}, \bibinfo{journal}{Science}
  \textbf{\bibinfo{volume}{289}}, \bibinfo{pages}{2323} (\bibinfo{year}{2000}).

\bibitem[{\citenamefont{D'Agosta et~al.}(2006)\citenamefont{D'Agosta, Sai, and
  {Di Ventra}}}]{dagosta06nl}
\bibinfo{author}{\bibfnamefont{R.}~\bibnamefont{D'Agosta}},
  \bibinfo{author}{\bibfnamefont{N.}~\bibnamefont{Sai}}, \bibnamefont{and}
  \bibinfo{author}{\bibfnamefont{M.}~\bibnamefont{{Di Ventra}}},
  \bibinfo{journal}{Nano Lett.} \textbf{\bibinfo{volume}{6}},
  \bibinfo{pages}{2935} (\bibinfo{year}{2006}).

\bibitem[{\citenamefont{Huang et~al.}(2006)\citenamefont{Huang, Xu, Chen, {Di
  Ventra}, and Tao}}]{huang06}
\bibinfo{author}{\bibfnamefont{Z.~F.} \bibnamefont{Huang}},
  \bibinfo{author}{\bibfnamefont{B.~Q.} \bibnamefont{Xu}},
  \bibinfo{author}{\bibfnamefont{Y.-C.} \bibnamefont{Chen}},
  \bibinfo{author}{\bibfnamefont{M.}~\bibnamefont{{Di Ventra}}},
  \bibnamefont{and} \bibinfo{author}{\bibfnamefont{N.~J.} \bibnamefont{Tao}},
  \bibinfo{journal}{Nano Lett.} \textbf{\bibinfo{volume}{6}},
  \bibinfo{pages}{1240} (\bibinfo{year}{2006}).

\bibitem[{\citenamefont{Kohn and Sham}(1965)}]{kohn65}
\bibinfo{author}{\bibfnamefont{W.}~\bibnamefont{Kohn}} \bibnamefont{and}
  \bibinfo{author}{\bibfnamefont{L.}~\bibnamefont{Sham}},
  \bibinfo{journal}{Phys. Rev.} \textbf{\bibinfo{volume}{140}},
  \bibinfo{pages}{A1133} (\bibinfo{year}{1965}).

\bibitem[{\citenamefont{Zangwill and Soven}(1980)}]{zangwill80}
\bibinfo{author}{\bibfnamefont{A.}~\bibnamefont{Zangwill}} \bibnamefont{and}
  \bibinfo{author}{\bibfnamefont{P.}~\bibnamefont{Soven}},
  \bibinfo{journal}{Phys. Rev. Lett.} \textbf{\bibinfo{volume}{45}},
  \bibinfo{pages}{204} (\bibinfo{year}{1980}).

\bibitem[{\citenamefont{Runge and Gross}(1984)}]{runge84}
\bibinfo{author}{\bibfnamefont{E.}~\bibnamefont{Runge}} \bibnamefont{and}
  \bibinfo{author}{\bibfnamefont{E.~K.~U.} \bibnamefont{Gross}},
  \bibinfo{journal}{Phys. Rev. Lett.} \textbf{\bibinfo{volume}{52}},
  \bibinfo{pages}{997} (\bibinfo{year}{1984}).

\bibitem[{\citenamefont{Ceperley and Alder}(1980)}]{ceperley80}
\bibinfo{author}{\bibfnamefont{D.~M.} \bibnamefont{Ceperley}} \bibnamefont{and}
  \bibinfo{author}{\bibfnamefont{B.~J.} \bibnamefont{Alder}},
  \bibinfo{journal}{Phys. Rev. Lett.} \textbf{\bibinfo{volume}{45}},
  \bibinfo{pages}{566} (\bibinfo{year}{1980}).

\bibitem[{\citenamefont{Perdew and Zunger}(1981)}]{perdew81}
\bibinfo{author}{\bibfnamefont{J.~P.} \bibnamefont{Perdew}} \bibnamefont{and}
  \bibinfo{author}{\bibfnamefont{A.}~\bibnamefont{Zunger}},
  \bibinfo{journal}{Phys. Rev. B} \textbf{\bibinfo{volume}{23}},
  \bibinfo{pages}{5048} (\bibinfo{year}{1981}).

\bibitem[{\citenamefont{Tal-Ezer and Kosloff}(1984)}]{tal-ezer84}
\bibinfo{author}{\bibfnamefont{H.}~\bibnamefont{Tal-Ezer}} \bibnamefont{and}
  \bibinfo{author}{\bibfnamefont{R.}~\bibnamefont{Kosloff}},
  \bibinfo{journal}{J. Chem. Phys.} \textbf{\bibinfo{volume}{81}},
  \bibinfo{pages}{3967} (\bibinfo{year}{1984}).

\end{thebibliography}

\end{document}